\documentstyle{europhys}

\def\And{{\rm and\ }}

\def\drm{{\rm d}}

\newif\ifboo \boofalse

\input{psfig.sty}

\newcommand{\paper}{Letter}

\psfigurepath{Figures}

\begin{document}
\euro{47}{4}{}{}
\Date{15 August 1999}
\shorttitle{Heating of trapped atoms near thermal surfaces}
\title{Heating of trapped atoms near thermal surfaces}
\author{C. Henkel\footnote{email: 
Carsten.Henkel@quantum.physik.uni-potsdam.de} 
\And M. Wilkens} 
\institute{
     Institut f\"ur Physik, Universit\"at Potsdam, Am Neuen Palais 10,
     14469 Potsdam, Germany}
%
%
\rec{8 February 1999}{17 June 1999}
%
%
%
\pacs{
\Pacs{03}{75}{Matter waves}
\Pacs{32}{80.Lg}{Mechanical effects of light on atoms and ions}
\Pacs{03}{67}{Quantum computation}
      }
\maketitle
\begin{abstract}
We study the electromagnetic coupling and concomitant heating
of a particle in a miniaturized trap close to a solid surface. 
Two dominant heating mechanisms are identified: 
proximity fields generated by thermally excited currents
in the absorbing solid and time-dependent image potentials due to
elastic surface distortions (Rayleigh phonons). 
Estimates for the lifetime of the trap ground state are given.
Ions are particularly sensitive to electric proximity fields:
for a silver substrate, we find a lifetime below one second 
at distances closer than some ten $\mu$m to the surface. Neutral
atoms may approach the surface more closely: if they have a magnetic moment,
a minimum distance of one $\mu$m is estimated in tight traps, 
the heat being transferred via magnetic proximity fields. 
For spinless atoms, heat is transferred by inelastic scattering
of virtual photons off surface phonons. The corresponding
lifetime, however, is estimated to be extremely long compared to
the timescale of typical experiments.
\end{abstract}
%
%
%
The last few years have witnessed an increasing interest
in tightly confining traps of cold particles. 
These devices allow to envisage a broad spectrum of applications 
ranging from 
single-mode coherent matter wave manipulation
and low-dimensional quantum gases 
\cite{Ovchinnikov97b,Hinds98,Mlynek98b,Haensch98}
to quantum logical registers \cite{Zoller95,Wineland98}.
Since steep trapping fields exist near surfaces, 
traps in their vicinity enjoy increasing popularity.
This raises the question at what timescale the
cold particles in these ``surface assisted traps''
will be heated up, and how they are coupled 
to the nearby bulk which is typically at room temperature
\cite{Wineland75,Lamoreaux97,James98}. 
The question is of primordial importance for the above-mentioned
applications since the heat transfer to the trap 
inevitably destroys the coherence of the matter waves \cite{Wineland98}.

In this \paper, we outline simple models that allow to compute
the lifetime of the trapped particle which is limited 
due to its coupling to thermal excitations 
of the nearby solid.
The interaction with thermal blackbody radiation is certainly
a candidate for a mechanism of heating and decoherence. 
Estimates given by Wineland and Dehmelt \cite{Wineland75}
and others \cite{Lamoreaux97,James98} show, however, that 
this source is negligible for typical trap configurations. 
This is mainly due to the fact that the trapped
particles are most sensitive to the field fluctuations at the resonant
trap oscillation frequency (a few MHz at most) 
which is rather low compared to thermal frequencies which are in the
THz range. More importantly, the resonant photon 
wavelengths are at least several meters. This
means that the particle is always located in the {\it near field\/} of 
its macroscopic environment where the electromagnetic field fluctuations
differ from the free-space blackbody field \cite{Agarwal75a,Greffet99}.
The excitations of the solid that give rise to this near-field effect
come in two species: fluctuating
electric currents related to the dissipation in the solid (finite
electric conductivity), and elastic waves (Rayleigh phonons)
that propagate along the surface. Current fluctuations generate 
electric and magnetic fields above the surface (``proximity fields'') 
that couple to the particle's charge, spin or polarizability.
Surface waves, on the other hand, distort the electrostatic
image of the particle in the solid and lead to a time-dependent 
image potential.
We find that ions are particularly sensitive to proximity fields
and estimate a typical lifetime of less than a second for distances
smaller than 10$\,\mu$m above a metal surface. Atoms, being neutral
particles, are less affected by the presence of the ``hot''
surface: for a nonzero magnetic moment, they survive several minutes
even at distances of a few micrometers. Finally, spinless atoms are
completely decoupled from the surface at experimentally relevant
time scales.

\section{Model}
\begin{figure}
\centerline{%
\psfig{figure=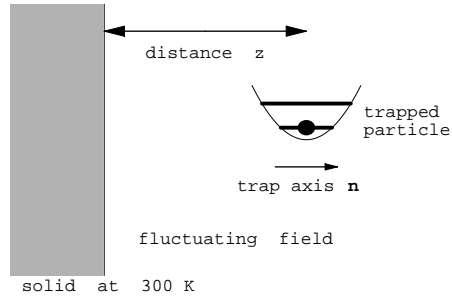,height=120pt,clip=}
}
\caption[]{Schematical setup of the experiment.}
\label{fig:trap-model}
\end{figure}
We consider a particle in the ground state $|0\rangle$ 
of a one-dimensional harmonic trap (oscillation frequency $\omega_{\rm t}$)
that is oriented along the unit vector ${\bf n}$
(see fig.\ref{fig:trap-model}) 
To simplify the calculations, we assume that the trap is located
above a flat surface whose distance $z$ from the trap center is large compared 
to the size $a = (\hbar/(2 M \omega_{\rm t}))^{1/2}$ 
of the ground state wave function ($M$ is the particle's mass). 
The interaction potential for the particle is then of the form
\begin{equation}
V(t) = - {\bf x} \cdot {\bf F}({\bf r}, t)
\label{eq:perturbation}
\end{equation}
where ${\bf x} = x {\bf n}$ is the displacement of the particle 
from the trap center and ${\bf F}( {\bf r}, t )$ is a fluctuating
force field at the trap position ${\bf r}$. 
If this force shows fluctuations at the trap frequency,
it may resonantly excite the particle to the first excited trap
state $|1\rangle$. According to second-order perturbation theory (Fermi's
Golden Rule), this happens with a rate
\begin{equation}
\Gamma_{1\gets 0}( {\bf r} ) = \hbar^{-2} 
\left| \langle 1 | x | 0 \rangle \right|^2  
\sum_{ij} n_i n_j S^{ij}_{F}( {\bf r}, {\bf r}; -\omega_{\rm t} )
\label{eq:2nd-order-rate}
\end{equation}
where $ \langle 1 | x | 0 \rangle$ is a ``dipole matrix element''
(equal to $a$ for a harmonic trap), and the cross correlation tensor
of the force fluctuations is defined by
\begin{equation}
S^{ij}_{F}( {\bf r}', {\bf r}; \omega )
= 
\int\limits_{-\infty}^{+\infty}\!\drm t\,
\langle
F_i( {\bf r}', t ) F_j( {\bf r}, 0 ) \rangle_T
\,
e^{{\rm i}\omega t }
,
\label{eq:def-spectrum}
\end{equation}
where $i,j$ denote cartesian indices. The average is taken in thermal 
equilibrium at temperature $T$. 
Note that the ``heating rate'' $\Gamma_{1\gets 0}$ 
also governs the decay of the coherence between different
trap levels, as is easily shown by deriving a master equation for the
particle's density matrix in the usual Born-Markov approximation.
Decoherence in ion traps was theoretically investigated by several
authors \cite{Wineland98,Wineland75,Lamoreaux97,James98,Milburn98,Knight98},
particularly in the wake of a recent experiment by Meekhof {\it et al.}\ 
\cite{Wineland96}.
In distinction to the present work, refs.\cite{Milburn98,Knight98}
focussed mainly on the decay of off-diagonal density matrix elements.

\section{Ion heating}
\begin{figure}
\centerline{%
\psfig{figure=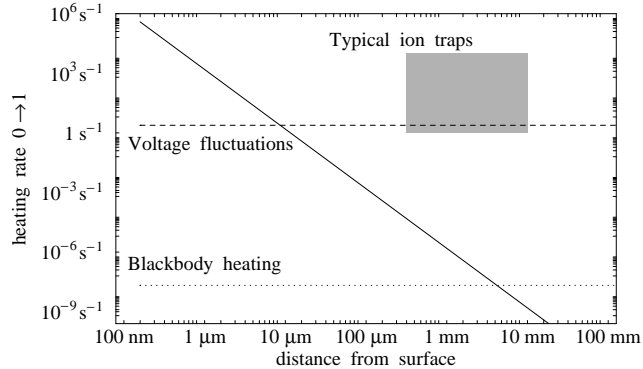,height=150pt,clip=}
}
\caption[]{Transition rate $0\to 1$ (`heating rate') for a 
trapped ion above a Ag surface, plotted vs.\ the trap
distance $z$. Trap oriented perpendicular to the surface 
(${\bf n} = {\bf e}_z$) with oscillation frequency $\omega_{\rm t}/2\pi
= 1\,$MHz.
Solid line: coupling to electric proximity fields
(ion mass $M = 40\,$amu, charge $q = e$, and temperature $T = 300\,$K). 
Dotted line: coupling to the blackbody field.
Dashed line: coupling to thermal voltage fluctuations in an ion trap,
as discussed in \cite{Wineland75,Lamoreaux97}
(endcap distance $1\,$mm, circuit resistance $R = 1\,\Omega$).
The shaded rectangle indicates the size and inverse lifetimes of
typical experiments \cite{Wineland96,Blatt99p}. 
}
\label{fig:ion-heating}
\end{figure}

The simplest case is that of a trapped ion. The force  
in~(\ref{eq:perturbation}) is given in terms of the electric field
as ${\bf F} = q {\bf E}$. To compute the electric field fluctuations, we
use the fluctuation--dissipation theorem that relates their spectral
density to the field's Green tensor $G_{ij}( {\bf r}', {\bf r}; \omega )$, 
{\it i.e.}, the field created at ${\bf r}'$ by an oscillating point dipole 
located at ${\bf r}$ \cite{Agarwal75a}:
\begin{equation}
S^{ij}_{E}( {\bf r}, {\bf r}; \omega ) = 
\frac{ 2 \hbar }{ 1 - e^{-\hbar\omega / T} }
\mathop{\rm Im}\, G_{ij}( {\bf r}, {\bf r}; \omega )
.
\label{eq:fluct-diss}
\end{equation}
(In our units, $k_{\rm B} = 1$.) The Green tensor contains a
free-space term $\mathop{\rm Im}\,G^{(bb)}_{ij} = 
\omega^3/ (6\pi\varepsilon_0c^3) \delta_{ij}$
that describes the black-body field, and a surface-dependent term
due to the reflection at the surface. For typical trap frequencies, the
electromagnetic wavelength is much larger than $z$ so that we may use
the quasi-static approximation for the Green tensor and get (in SI units)
\begin{equation}
\mathop{\rm Im}\, G^{\rm (s)}_{ij}( {\bf r}, {\bf r}; \omega ) =
\frac{ s_{ij} }{ 16\pi\varepsilon_0\, z^3 }
\mathop{\rm Im}\,\frac{\epsilon(\omega) - 1}{\epsilon(\omega) + 1}
\label{eq:dipole-field}
\end{equation}
where $s_{ij}$ is a diagonal tensor with $s_{xx} = s_{yy} = \frac12$,
$s_{zz} = 1$.
A reasonable model for the dielectric function $\varepsilon(\omega)$
in~(\ref{eq:dipole-field}) is that of a Drude metal \cite{Kittel}.
For frequencies below the electronic damping rate, the electrostatic
reflection coefficient in~(\ref{eq:dipole-field}) becomes
\begin{equation}
\mathop{\rm Im}\,\frac{\epsilon(\omega) - 1}{\epsilon(\omega) + 1}
=
2 \omega \varrho \varepsilon_0 
\label{eq:e-static-r}
\end{equation}
where $\varrho$ is the metal's specific resistance at the trap
frequency. Finally, we may also take the high-temperature limit of the
Bose-Einstein factor in~(\ref{eq:fluct-diss}).
This gives a surface-induced heating rate for the trapped ion that follows
a power law as a function of distance $z$
\begin{equation}
\Gamma^{(s)}_{1\gets 0}( z ) =
\frac{ q^2 T \varrho }{ \hbar \omega_{\rm t} M }
\frac{ 1 + n_z^2 }{ 16\pi z^3 }
.
\label{eq:ion-heating}
\end{equation}
Typical values are shown in fig.\ref{fig:ion-heating} where we see that
close to a silver surface at room temperature, the ion's lifetime is shorter 
than $1\,$s as soon as the trap gets closer than about 10$\,\mu$m.
The figure also shows that close to the metal surface, the electric
proximity field dominates over thermal blackbody radiation (the dotted 
line). We note that this result may be understood easily using
the analogy between the trapped ion 
and an oscillating electric dipole that is damped both radiatively
and nonradiatively. (This analogy has already been exploited to
interpret decoherence times in early ion trapping experiments,
see ref.\cite{Wineland75}.)
Recall that at zero temperature, an excited oscillator decays to its
ground state because of photon emission or nonradiative energy transfer
into the environment (here, the absorbing metal surface) \cite{Chance78}. 
For distances well below the transition wavelength, the decay is dominated 
by nonradiative transfer from the dipole's near field to the metal, as
described by eq.(\ref{eq:dipole-field}). At low frequencies
or, equivalently, high temperatures, the decay rate $1\to0$ is
dominated by stimulated emission and hence proportional to the
temperature, see eq.(\ref{eq:ion-heating}). On the other hand,
the principle of detailed balance implies that the inverse 
transition $0\to1$ that we are presently interested in, 
occurs at the same rate in this regime. The ion is hence heated up
because thermal energy is transferred from the metal surface to the
ion's near field in a nonradiative way. (A similar reasoning has been
presented by Anglin, Paz, and Zurek \cite{Anglin97}, though
they focussed on the energy transfer from a fast-moving ion into 
a solid and the concomitant decoherence.) 

We may also compare the heating rate~(\ref{eq:ion-heating})
to the calculations of Wineland and Dehmelt \cite{Wineland75},
Lamoreaux \cite{Lamoreaux97}, and James \cite{James98}
who studied the heating due to 
thermal voltage fluctuations across the endcaps of a Paul trap.
These authors' results are recovered
(up to a geometrical factor of order unity) if we replace 
in~(\ref{eq:ion-heating}) the quantity $\varrho / z$ with the electric
resistance $R$ of the endcap circuitry, and interpret $z$ as the endcap
distance: the quantity $R T / z^2$ then gives the power spectrum of the
thermal electric field between the endcaps.
The present model shows, however, that even in front of a single surface,
electric proximity fields leak out of the solid whose power spectrum
increases even stronger with decreasing distance.

\section{Magnetic proximity fields}
\begin{figure}
\centerline{%
\psfig{figure=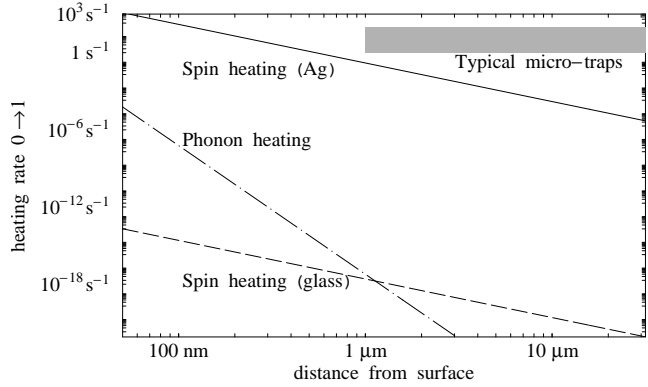,height=150pt,clip=}
}
\caption[]{Transition rate $0\to 1$ (`heating rate') vs.\ trap distance $z$
for a trapped particle with mass $M = 40\,$amu. The trap is
oriented perpendicular to the surface and has frequency
$\omega_{\rm t}/2\pi = 100\,$kHz.
Solid line: coupling to magnetic proximity fields above a silver substrate
(magnetic moment $\mu/(2\pi\hbar) = 1.4\,$MHz/G\ =\ 1 Bohr magneton,
spin $\frac12$, specific resistance $\varrho = 1.6\times10^{-6}\,\Omega$cm
\cite{Kittel}).
Dashed line: same above a glass substrate ($\varrho = 10^{11}\,\Omega$cm).
Dash-dotted line: coupling to surface phonons via the van-der-Waals
interaction ($c_3/(2\pi\hbar) = 1\,$kHz$\,\mu$m$^3$,
Ag substrate with $\hbar\omega_{\rm D} = 225\,$K, $M_{\rm s} = 108\,$amu,
$\eta = 0.75$ \cite{Desjonqueres}). 
The heating rate due to blackbody magnetic fields (not shown) is of
the order of $10^{-39}\,{\rm s}^{-1}$.
}
\label{fig:Zeeman-heating}
\label{fig:phonon-heating}
\end{figure}

Trapped ions and atoms frequently have a magnetic moment {\boldmath $\mu$}
and are therefore heated by time-dependent magnetic fields. The force
derives from the Zeeman interaction
\begin{equation}
V( {\bf r}, t ) = - \mbox{\boldmath $\mu$}\cdot{\bf B}( {\bf r}, t )
.
\label{eq:Zeeman-potential}
\end{equation}
As mentioned in the introduction, magnetic fields are created by
fluctuating currents in the solid. Following the seminal work
of Lifshitz \cite{Lifshitz56}, the spectral density of these currents
is proportional to the imaginary part of the dielectric function
$\varepsilon( {\bf r}; \omega )$. It has recently
been proven that the introduction of fluctuating currents
also provides a consistent framework
to quantize the electromagnetic field in absorbing and dispersive  
dielectrics \cite{Barnett91,Knoell98b}. 
We use the following representation of the current Fourier
transform ${\bf j}( {\bf r}; \omega )$ in terms of spatially uncorrelated
boson operators ${\bf f}( {\bf r}; \omega )$ \cite{Knoell98b}
\begin{equation}
{\bf j}( {\bf r}; \omega )
=
\omega \sqrt{ 2 \hbar \varepsilon_0 
\mathop{\rm Im}\,\varepsilon( {\bf r}, \omega )
}
\,{\bf f}( {\bf r}; \omega )
.
\label{eq:current-spectrum}
\end{equation}
For a nonmagnetic solid filling the half-space $z<0$, the Biot-Savart
law yields the following magnetic field cross correlation tensor 
on the vacuum side $z', z > 0$
\begin{equation}
S_B^{ij}( {\bf R}, z', {\bf R}, z; \omega )
=
\frac{ \mu_0^2 \omega^2 \hbar \varepsilon_0
\mathop{\rm Im}\, \varepsilon( \omega ) }{ 1 - e^{- \hbar \omega / T} }
\frac{ t_{ij} }{ 8 \pi (z' + z) }
\label{eq:magnetic-spectrum}
\end{equation}
where ${\bf R} = (x, y)$ denotes coordinates parallel to the surface
and $t_{ij}$ is a diagonal tensor with $t_{xx} = t_{yy} = \frac32$
and $t_{zz} = 1$. For simplicity, we focus on a trap oriented perpendicular
to the surface. Taking gradients with respect to $z$ and $z'$
and putting $z' = z$, we get the desired spectral density of the 
($z$-component of the) magnetic force. Using again the
low-frequency limit for the dielectric function,
$\mathop{\rm Im}\, \varepsilon( \omega ) = 1 / (\varepsilon_0 \omega
\varrho)$, and the high-$T$
limit of the Bose-Einstein factor, we end up with the following heating rate
\begin{equation}
\Gamma_{1\gets 0} = 
\frac{ \mu_0^2 T }{ \hbar \omega_{\rm t} M \varrho }
\frac{  
\left\langle 3 \mbox{\boldmath$\mu$}^2 - \mu_z^2 
\right\rangle
}{ 128 \pi z^3 } 
\label{eq:Zeeman-heating}
\end{equation}
This quantity is represented in fig.\ref{fig:Zeeman-heating} for a trap
close to an Ag substrate, and one observes a relatively large heating rate 
of the order of $10^{-2}\,$s$^{-1}$ at a distance of $1\,\mu$m. 
A glass substrate gives a much smaller heating rate (dashed line)
because its resistance is larger. 
We conclude that one has to avoid either metal surfaces or 
particles with spin if one wants to store atoms coherently 
over timescales longer than, say, a few minutes. 

\section{Heating of neutral, spinless particles}
We now turn to the heating of a neutral, spinless particle. 
One would expect it to be less sensitive to fluctuating electric fields,
although these couple to the atomic polarizability $\alpha$ via the
Stark shift $V = \frac{ \alpha }{ 2 } {\bf E}^2$.
We have computed the corresponding heating rate along similar lines
as for the trapped ion and found extremely small values (below
$10^{-12}\,$s$^{-1}$ even at distances around $100\,$nm above a Ag substrate). 
In fact, atom heating is dominated by a different effect: the distortion
of the surface by thermal oscillations leading to a time-dependent
image potential.
The corresponding force may be easily computed using the following 
effective interaction (strictly valid for a rarefied solid and in the 
quasi-static limit): 
we integrate a $1/r^6$ dipole-dipole interaction between the atom and 
the half-space filled by the solid. For a flat interface, one finds the
usual van-der-Waals potential $-c_3/z^3$. If the interface is corrugated,
but with an amplitude small compared to the distance $z$, one gets
a first-order correction $V^{(1)}( {\bf r}, t )$ of the form
\cite{Henkel98b}
\begin{equation}
V^{(1)}( {\bf r}, t ) = \sum_{\bf Q}
g( {\bf Q}; z ) u_{\bf Q}( t ) \exp{{\rm i} {\bf Q}\cdot{\bf R}}
.
\label{eq:vdW-force}
\end{equation}
In this expression,
${\bf Q}$ is a two-dimensional `lateral' wave vector (in the $xy$-plane
parallel to the non-excited surface),
${\bf R}$ are the lateral coordinates of the trap center, 
the $u_{\bf Q}( t )$ are the elastic displacements of the surface
at wave vector ${\bf Q}$,
and, finally, the coupling
coefficients are given by $g( {\bf Q}; z ) = - (3 c_3 Q^2 / (2 z^2))
K_2( Q z )$ with $K_2$ a modified Bessel function. 

For simplicity, we restrict ourselves to Rayleigh waves 
\cite{Desjonqueres} whose displacement amplitudes are confined 
to the vicinity of the surface. 
Their amplitudes may be written in terms of boson operators
\begin{equation}
u_{\bf Q}( t ) = 
\sqrt{ \frac{ \hbar \eta \pi }{ N_{\rm s} M_{\rm s} \omega_{\rm D} } }
\left(
a_{\bf Q}( t ) + a_{-\bf Q}^\dag( t )
\right)
,
\label{eq:Rayleigh-wave}
\end{equation}
where the dimensionless parameter $\eta$ specifies the decay of the 
surface wave into the bulk (it depends on the ratio between the 
bulk and Rayleigh wave sound velocities), $N_{\rm s}$
is the number of surface atoms per quantization area, $M_{\rm s}$ is their
mass, and $\omega_{\rm D}$ is the Debye frequency \cite{Desjonqueres}. 
Using thermal
Bose-Einstein statistics for the Rayleigh waves, we can compute
the spectrum of the surface oscillations at the trap frequency. 
In the result, the magnitude of the Rayleigh wave vector is fixed
to $Q = \omega_{\rm t} / v_{\rm R}$ where $v_{\rm R}$ is the Rayleigh
wave sound velocity. Typically, the sound wavelength 
is much longer than the trap distance, and 
the coupling constant $g( {\bf Q}; z)$ may thus be approximated 
by its asymptotic form for $Q z \ll 1$. The heating rate then equals
\begin{equation}
\Gamma_{1\gets 0} =
\frac{ T c_3^2 }{ 
\hbar\omega_{\rm t} \omega_{\rm D}^3 M M_{\rm s} }
\frac{ 72 \pi^3 \eta(1 + \eta^2) }{ z^{10} }
.
\label{eq:phonon-heating}
\end{equation}
As shown in fig.\ref{fig:phonon-heating}, reasonable parameters give 
a still very small heating rate 
(below $10^{-6}\,$s$^{-1}$ at $100\,$nm distance). Even on a timescale
of hours, the spinless atom is thus decoupled from the thermal excitations 
of the solid. We note that in evanescent field traps, the interaction
with light scattered off surface/bulk impurities may be a dominant heating
mechanism.%
\footnote{C. Henkel, R. Adams, H. Gauck, D. Schneble, T. Pfau, C. I. Westbrook,
and J. Mlynek, in preparation.}.

\section{Conclusion}
In the vicinity of a solid surface at room temperature, fluctuating
electric and magnetic fields couple to trapped particles and
induce a finite lifetime of the trap's ground state. If one wants
a lifetime longer than about one second, ions must not 
be closer to the surface than about some tens of micrometers.
This means that miniaturized coherent ion traps are difficult to 
realize at room temperature.
Magnetic proximity fields are weaker, and atoms with spin live for many seconds
in the ground state even at distances of a micron. The best candidates
for long-time storage are spinless atoms because they are nearly insensitive 
to stray fields. Their ground state lifetime is much longer than hours
and is mainly limited by surface waves that distort the electrostatic image 
potential.

The preceding estimates have been obtained from simple models for the
trap geometry, neglecting the finite height of the trap potential and 
assuming a homogeneous substrate. 
An obvious extension would be to allow for layered media. Another
point is the inclusion of a cross-coupling between fluctuating 
and static trapping fields. This increases the heating due
to the Stark potential, {\it e.g.}, because the atom acquires
a static electric dipole moment in the trap. Finally, in some traps
(magnetic or near-resonant optical) the coupling to non-trapped
internal states of the particle, {\it e.g.}, hyperfine or magnetic states,
may lead to a relevant loss rate.

\section{Acknowledgements}
C.H. is a research fellow of the Deutsche For\-schungs\-ge\-mein\-schaft.

%
%
%
%


\vskip-12pt
\begin{thebibliography}{10}

\bibitem{Ovchinnikov97b}
  Y.~B. Ovchinnikov, I. Manek, and R. Grimm, Phys. Rev. Lett. {\bf 79},  2225  
  (1997).

\bibitem{Hinds98}
E.~A. Hinds, M.~G. Boshier, and I.~G. Hughes, Phys. Rev. Lett. {\bf 80},  645
  (1998).

\bibitem{Mlynek98b}
H. Gauck, M. Hartl, D. Schneble, H. Schnitzler, T. Pfau, and J. Mlynek, Phys. Rev. Lett. {\bf 81},  5298  (1998).

\bibitem{Haensch98}
J. Fortagh, A. Grossmann, C. Zimmermann, and T.~W. H\"ansch, Phys. Rev. Lett.
  {\bf 81},  5310  (1998).

\bibitem{Zoller95}
J.~I. Cirac and P. Zoller, Phys. Rev. Lett. {\bf 74},  4091  (1995).

\bibitem{Wineland98}
C. Monroe, W.~M. Itano, D. Leibfried, B.~E. King, 
  and D.~M. Meekhof, J. Res. Natl. Inst. Stand. Technol. {\bf 103},
  259  (1998).

\bibitem{Wineland75}
D.~J. Wineland and H.~G. Dehmelt, J. Appl. Phys. {\bf 46}, 919 (1975).

\bibitem{Lamoreaux97}
S.~K. Lamoreaux, Phys. Rev. A {\bf 56},  4970  (1997).

\bibitem{James98}
D.~F.~V. James, Phys. Rev. Lett. {\bf 81},  317  (1998).

\bibitem{Agarwal75a}
G.~S. Agarwal, Phys. Rev. A {\bf 11},  230  (1975).

\bibitem{Greffet99}
R. Carminati and J.-J. Greffet, Phys. Rev. Lett. {\bf 82},  1660  (1999).

\bibitem{Milburn98}
S. Schneider and G.~J. Milburn, Phys. Rev. A {\bf 57},  3748  (1998).

\bibitem{Knight98}
M. Murao and P. Knight, Phys. Rev. A {\bf 58},  663  (1998).

\bibitem{Wineland96}
D.~M. Meekhof, C. Monroe, B.~E. King, W.~M. Itano, and D.~J. Wineland, 
  Phys. Rev. Lett. {\bf 76},  1796  (1996); 
  {\it ibid.}\ {\bf 77}, 2346(E) (1996).

\bibitem{Kittel}
C. Kittel, {\em Introduction to Solid State Physics}, 5th ed. (Wiley, New York,
  1976).

\bibitem{Chance78}
R.~R. Chance, A. Prock, and R. Silbey,  in {\em Advances in Chemical Physics
  XXXVII}, edited by I. Prigogine and S.~A. Rice (Wiley \& Sons, New York,
  1978).

\bibitem{Anglin97}
J.~R. Anglin, J.~P. Paz, and W.~H. Zurek, Phys. Rev. A {\bf 55},  4041  (1997).

\bibitem{Blatt99p}
F. Schmidt-Kaler, E. Peik, personal communications (1999).



\bibitem{Lifshitz56}
E.~M. Lifshitz, Soviet Phys. JETP {\bf 2},  73  (1956) [{\em J. Exper.
  Theoret. Phys. USSR} {\bf 29}, 94 (1955)].

\bibitem{Barnett91}
B. Huttner, J.~J. Baumberg, and S.~M. Barnett, Europhys. Lett. {\bf 16},  177
  (1991).

\bibitem{Knoell98b}
S. Scheel, L. Kn\"{o}ll, and D.-G. Welsch, Phys. Rev. A {\bf 58},  700  (1998).

\bibitem{Desjonqueres}
M.-C. Desjonqu\`eres and D. Spanjaard, {\em Concepts in Surface Physics}, 2nd
  ed. (Springer, Berlin, 1996).

\bibitem{Henkel98b}
C. Henkel and V. Sandoghdar, Opt. Commun. {\bf 158},  250  (1998).

\end{thebibliography}

%

\end{document}